\documentstyle[emulapj]{article}

\begin{document}

\lefthead{MAGNETIC PROPELLER OUTFLOWS}
\righthead{LOVECALE, ROMANOVA, BISNOVATYI--KOGAN}

\submitted{}

\title{MAGNETIC PROPELLER OUTFLOWS}

\author{R.V.E. Lovelace}
\affil{Department of Astronomy,
Cornell University, Ithaca, NY 14853-6801;
rvl1@cornell.edu }
\author{M.M. Romanova}
\affil{Space Research Institute,
Russian Academy of Sciences, Moscow, Russia; and\\
Department of Astronomy,
Cornell University, Ithaca, NY 14853-6801;
romanova@astrosun.tn.cornell.edu}
\author{G.S. Bisnovatyi-Kogan}
\affil{Space Research Institute,
Russian Academy of Sciences, Moscow,
Russia; gkogan@mx.iki.rssi.ru}

\medskip

\slugcomment{Accepted to the Astrophysical Journal}

\begin{abstract}
 A model is developed for magnetic
`propeller'-driven outflows
which cause a rapidly rotating
magnetized star accreting from a
disk to spin-down.
Energy and angular
momentum lost by the star goes
into expelling most of the accreting
disk matter.  The theory gives
an expression for the effective Alfv\'en
radius $R_A$ (where the inflowing
matter is effectively stopped)
which depends on the mass
accretion rate, the star's mass and
magnetic moment, {\it and} the star's
rotation rate.
 The model points to a mechanism
for `jumps' between spin-down and
spin-up evolution and for the reverse
transition, which are changes between
two possible equilibrium configurations
of the system.
 In for example the transistion from
spin-down to spin-up states
the Alfv\'en radius $R_A$ decreases
from a value larger than
the corotation radius to one which
is smaller.  In this transistion
the `propeller' goes from being ``on''
to being ``off.''
 The ratio of the spin-down
to spin-up torque (or the ratio
for the reverse change) in a jump
is shown to be of order unity.

\end{abstract}


\begin{figure*}[t]
\epsscale{0.5}
\plotone{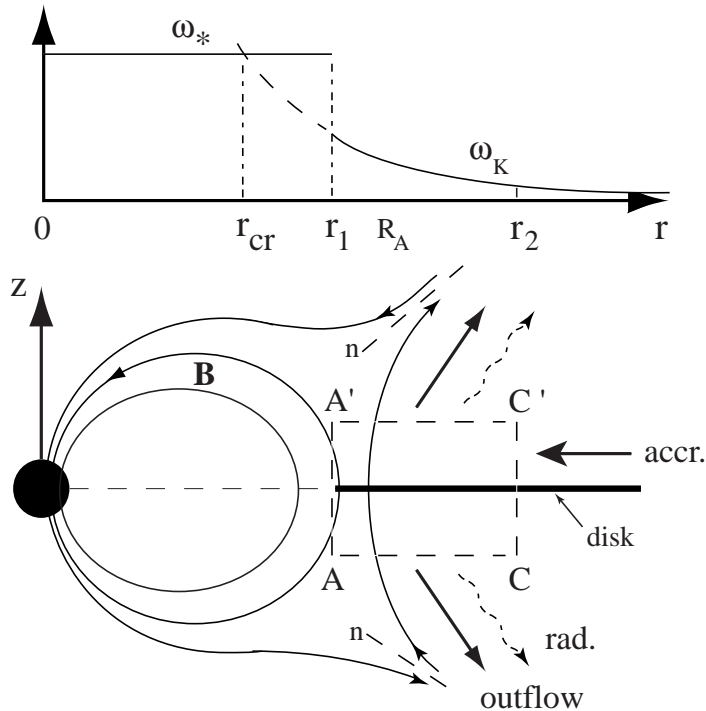}
\caption{
Geometry of disk accretion
to a rapidly rotating star with an aligned
dipole magnetic field.
  Here, $r_1$
and $r_2$ are the boundaries of the
region considered in the text; $R_A$
is the effective radius where the
outflow begins;  $\omega_*$
is the star's rotation rate;  $\omega_K$
is the Keplerian rotation rate of the
accretion disk; and $r_{cr} \equiv
(GM/\omega_*^2)^{1/ 3}$
is the corotation radius.
  The magnetic field in the vicinity of $r_1$
has an essential time-dependence owing to the
continual processes of stellar flux leaking
outward into the disk, the resulting field
loops being inflated by the differential
rotation (LRBK), and the reconnection between
the open disk field and the closed stellar
field loops.
 The dashed lines marked by the letters
`n'  indicate neutral surfaces along which
reconnection occurs.
}
\label{fig1}
\end{figure*}

\section{Introduction}

Observations of some
X-ray pulsars show remarkable `jumps' between
states where the pulsar is spin-ning-down
to one where it is spinning-up.
 Examples include the
objects Cen X-3 (Chakrabarty et al. 1993) and
GX 1+4 (Chakrabarty et al. 1997; Cui 1997).
 The theoretical problem of disk accretion
to a rotating magnetized star has been
discussed in many works over a long
period (Pringle \& Rees 1972; Lynden-Bell
\& Pringle 1974; Ghosh \& Lamb 1979;
Wang 1979;
Lipunov 1993;
Shu
et al. 1994; Lovelace, Romanova, \&
Bisnovatyi-Kogan 1995 (hereafter LRBK);
Li \& Wickramasinghe 1997).
  However, except for the
work by Li \& Wickramasinghe
(1997), the studies do not specifically
address the `propeller' regime (Illarionov
\& Sunyaev 1975) where the
rapid rotation of the star's magnetosphere
acts to expell most of the accreting matter
and where the star spins-down.
   Recent computer simulation studies
of disk accretion to a rotating star
with an aligned dipole magnetic field
(Hayashi, Shibata, \& Matsumoto 1996; Goodson,
Winglee, \& B\"ohm 1997; Miller \& Stone 1997)
provide evidence of time-dependent outflows
but do not give definite evidence for
a `propeller' regime with spin-down of the star.
 The present work considers the `propeller' regime
and develops a
simple  physical model
where the energy and angular momentum
lost by the rotating star goes into
a magnetically driven outflow.

\section{Theory}

 We consider the problem of disk accretion
onto a rotating magnetized star which
has an aligned dipole
magnetic field.
 We focus on the limit
where the star is rotating rapidly and
the disk-star
 configuration is as
sketched in Figure 1 (see Figure 3 of
LRBK).
We consider the flow of mass, angular
momentum, and energy into and out
of the annular region indicated
by the box $A'ACC'$ in this figure,
where $A'A$ is at radius $r_1$ and
$CC'$ is at $r_2$.   Notice that
at this point the values of $r_1$
and $r_2$ are {\it unknown}.   They
are determined by the physical
considerations discussed here.

\placefigure{fig1}

Consider first the outer surface
$CC'$ through the disk.  The influx
of mass into the considered region
is
$$\dot{M}_{accr}=2\pi r_2\int_{-h}^{h}
dz~[\rho~ (-v_r)]_{2}~>0~,
\eqno(1)
$$
where $h$ is the half-thickness of the
disk, and the $2$ subscript
indicates evaluation at $r=r_2$.
We assume that mass accretion
rate for $r \geq r_2$ is approximately
constant equal to $\dot M_{accr}$.
That is, we consider that outflow
from the disk is negligible for $r\geq r_2$.
The influx of angular momentum into
the considered region is
$$
\dot{{L}}_{accr} = \dot{M}_{accr}(rv_\phi)_2
-T_2~.
\eqno(2a)
$$
Here,
$$ T_2= 2\pi r_2^2
\int_{-h}^h dz~ (T_{r\phi }^{vis})_2~,
\eqno(2b)
$$
where $T^{vis}_{r\phi }$ is the viscous
contribution to the stress tensor
which includes {\it both} the turbulent
hydrodynamic and turbulent magnetic stresses.
The influx of energy into the considered region
is
$$
\dot{E}_{accr} = \dot{M}_{accr}
\left(-{ GM \over 2 r_2}+w_2\right)
- \omega_2 T_2 ~,
\eqno(3)
$$
where $\omega \equiv v_\phi/r
\approx v_K/r$ with $v_K\equiv
\sqrt{GM/r}$ the Keplerian speed and $M$
the mass of the star, and where $w_2$ is
the enthalpy.  For conditions of interest here the
disk at $r_2$ is geometrically thin
so that $w \sim c_s^2 \ll v_K^2$, where
$c_s$ is the sound speed.

Consider next the fluxes of mass,
angular momentum, and energy across the surface
$AA'$ in Figure 1.
 For the physical regime considered,
where $v_K(r_1)/r_1 \ll \omega_*$ ($=$ the angular
rotation rate of the star), the
mass accretion across the $AA'$ surface
is assumed to be small
compared with $\dot{M}_{accr}$.
 The reason for this is that any plasma
which crosses the $AA'$ surface will
be `spun-up' to an angular velocity
$\omega_*$ (by the magnetic force)
which is substantially larger than
the Keplerian value, and thus it
will be thrown outwards.
Thus the efflux of angular
momentum across this surface
from the considered region is
$\dot{L}_1= -T_1~.
$
The efflux of energy across this surface is
$\dot{E}_1 = -\omega_*T_1,
$
where $\omega_*$ is the angular rotation rate of
the star and the inner magnetosphere as shown
in Figure 1.  For the conditions considered, the
star slows down and loses rotational energy so
that $T_1 >0$.

 We have
${\dot E_1 / \dot{L}_1} = \omega_*$.
Because the interaction of the star with the
accretion flow is by assumption entirely
across the surface $AA'$,
this is consistent with the spin-down of
a star with constant moment of inertia $I$;
that is,
$ {\Delta E_* /\Delta {L}_*}
={I_* \omega_* \Delta \omega_*/
(I_* \Delta \omega_*)} = \omega_*$.

Next we consider the mass, angular momentum,
and energy fluxes across the surfaces $A'C'$
and $AC$ in Figure 1.
As mentioned,
accretion to the star is
small for $r_{cr} \ll r_1$ where
$r_{cr}$ is the corotation radius as
indicated in Figure 1.
Thus, the mass accretion goes mainly into
outflows,
$\dot{M}_{out} \approx \dot{M}_{accr},
$
where ``$out$'' stands for outflows.
  The angular momentum outflow
across the surfaces $A'C'$ and
$AC$,
$\dot{L}_{out}$,
must be the difference between the angular momentum
lost by the star and the incoming angular momentum
of the accretion flow.   The angular momentum
carried by radiation from the disk is negligible
because $(v_K/c)^2 \ll 1$, where $c$ is the speed
of light.
That is,
$
\dot{L}_{out} =\dot{L}_{accr}
-\dot{L}_1.
$
The energy outflow across the $A'C'$ and $AC$ surfaces
is
$\dot  E_{out} +\dot{E}_{rad}=
\dot{E}_{accr}-\dot{E}_1,
$
where $\dot{E}_{rad}$ is the radiation energy loss
rate from the disk surfaces between $r=r_1$ and $r_2$,
and $\dot{E}_{out}$ is the rate of energy loss
carried by the outflows.

 Angular momentum conservation gives
$$\dot{L}_{out} =\dot{M}_{accr} (r v_\phi)_2 -
T_2 +T_1~.
\eqno(4)
$$
We have
$$\dot  E_{out} +\dot{E}_{rad}=
-{GM \dot{M}_{accr} \over 2 r_2}
-\omega_2 T_2 +\omega_* T_1~.
\eqno(5)
$$
We can solve equation (4) for $T_2$
and thereby eliminate this quantity from the energy
equation (5).  This gives
$$
\dot  E_{out} +\dot{E}_{rad}=
-{3G M \dot{M}_{accr} \over 2 r_2} + \omega_2 \dot
{L}_{out} +(\omega_* -\omega_2)T_1~.
\eqno(6)
$$

The preceeding equations are independent
of the nature of the outflows from the disk.
At this point we consider the case of
magnetically driven outflows as treated
by Lovelace, Berk, and Contopoulos (1991,
hereafter LBC).
  In the LBC model the outflows
come predominantly from an annular
inner region of the disk of radius $\sim R_A$
where the disk rotation rate is
$\omega_0 = \sqrt{GM/R_A^3}$.
  Thus we assume that the outflows
come from a region of the disk which
is approximately in Keplerian rotation.
 For the present situation, shown in
Figure 1, it is clear that we must have
$
r_1 <R_A <r_2~.
$
Further, we will assume $R_A$ is close in
value to $r_1$ with $(R_A -r_1)/R_A
{\buildrel < \over \sim 1}$.
For conditions where the outflow from the
disk is relatively low temperature (sound
speed much less than Keplerian speed),
equations (16) and (18) of LBC imply
the general relation
$
\dot{E}_{out} = \omega_0~\dot{L}_{out}-
{3GM\dot M_{out}/( 2 R_A)}.
$
This equation can be used to
eliminate $\dot{L}_{out}$
in favor of $\dot{E}_{out}$ in equation (6).
Recalling that $\dot{M}_{out} =\dot{M}_{accr}$
we have
$$
(1-\delta^{3\over2})
\dot{E}_{out} +\dot{E}_{rad}=
$$
$$
(\omega_*-\omega_2) T_1
-{3GM\dot{M}_{accr} \over 2 r_2}
(1-\delta^{1\over2})~,
\eqno(7)
$$
where $\delta \equiv R_A/r_2 <1$.

The energy dissipation in the
region of the disk $r=r_1$ to $r_2$
heats the disk and this heat energy
is transformed into outgoing radiation
$\dot{E}_{rad}$.
 Thus we have
$$
\dot{E}_{rad} ={3\over 2}\int_{r_1}^{r_2} dr~
{G M\dot{M}_{disk}(r)
\over r^2}~,
\eqno(8)
$$
(Shakura 1973; Shakura \& Sunyaev 1973),
where $\dot{M}_{disk}(r_2) =\dot{M}_{accr}$
and $\dot{M}_{disk}(r_1) \approx 0$.
The essential change in $\dot{M}_{disk}(r)$ occurs
in the vicinity of $R_A$ so that
$
\dot{E}_{rad} \approx (3 G M \dot{M}_{accr}/
2)({R_A^{-1}} - {r_2^{-1}} ).
$
Thus equation (7) becomes
$$
(1-\delta^{3\over2}) \dot{E}_{out}=
\left(\omega_*-\omega_2 \right)T_1
-{3GM\dot{M}_{accr} \over 2 R_A}(1-\delta^{3\over
2})~.
\eqno(9)
$$
Thus the power from the spin-down of the star
$\omega_*T_1$ must be larger than a certain
value in order to drive the outflow.

  For magnetically driven outflows, the
value of $\dot{E}_{out}$ can be written as
$$
\dot{E}_{out} = {3\over 2} ~{\cal F}_0^{2\over3}~
\omega_0^{4\over 3}~ R_A^{8\over 3} ~B_0^{4\over 3}~
\dot{M}^{1\over 3}_{accr}~,
\eqno(10)
$$
(equation 34 of LBC),
where ${\cal F}_0$ is a dimensionless numerical
constant $\buildrel >\over \sim 0.234$, and $B_0$ is
the poloidal magnetic field at the base of the outflow
at $r=R_A$.   We take the simple estimate
$
B_0= {\mu / R_A^3}~,
$
which omits corrections for example for compression of the
star's field by the inflowing plasma.

\begin{figure*}[t]
\epsscale{0.5}
\plotone{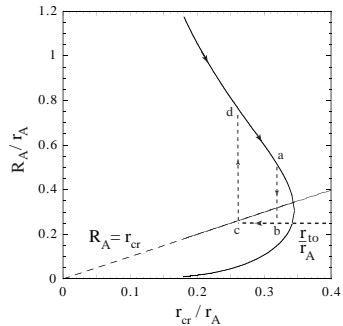}
\caption{
The plot shows
the {\it effective} Alfv\'en radius
$R_A$ (normalized by the {\it nominal}
Alfv\'en radius $r_A$)
as a function of the normalized
corotation radius $r_{cr}/r_A$ for
$\bar{\alpha}=0.2$ obtained
from equation (12).
 We have
neglected the $\delta^{3/2}=(R_A/r_2)^{3/2}$
terms compared
with unity and taken the value ${\cal F}_0
=0.234$ from LBC.
 Only the part of the solid curve above the
dashed line  $R_A=r_{cr}$
is consistent with our assumptions.
The dashed horizontal line $r_{to}$ indicates
the ``turnover radius'' of the disk rotation
curve or effective Alfv\'en radius in the regime
discussed by LRBK where $\dot{M}_{accr}$ falls onto
the star and the star spins-up.
 For this line the turbulent magnetic diffusivity
of the disk is taken to be $\alpha D = 0.1$.
The points $a~b~c~d$ and the associated
vertical lines represent possible transitions
between spin-down and spin-up of the pulsar as
discussed in the text.
}
\label{fig2}
\end{figure*}

   Next we consider the torque on the
star $T_1$.
   Because most of the matter
inflowing in the accretion disk at $r=r_2$ is driven off
in outflows at distances $r>r_1$, the  stress
is necessarily due to the magnetic field.
  The magnetic field in the vicinity of $r_1$
has an essential time-dependence owing to the
continual processes of stellar flux leaking
outward into the disk, the resulting field
loops being inflated by the differential
rotation (LRBK), and the reconnection between
the open disk field and the closed stellar
field loops.
  The time scale of these
processes is
$t_1 ~{\buildrel < \over \sim}~ 2\pi r_1/v_K(r_1)$.
We make
the estimate of the torque
$
T_1 = -2\pi r_1^2
(2 \Delta z )
{\langle B_{r} B_{\phi}\rangle_1 /( 4 \pi)},
$
where $\Delta z$ is the vertical half-thickness of
the region where the magnetic stress is significant,
and where the angular brackets denote a time average
of the field quantites at $r \sim r_1$.
   The magnetic field components, $B_r,~B_\phi$, with
$B_\phi \propto - B_r$ necessarily, must be of
magnitude less than or of the
order of the dipole field
$B_1=\mu/r_1^3$ at $r=r_1$.
  The fact that $B_\phi$ has the opposite sign
to that of $B_r$ is due to the fact that
the $B_\phi$ field arises differential
rotation between the region $r<r_1$, which
rotates at rate $\omega_*$, and the region
$r>r_1$, which rotates at
rate $\omega_K(r_1) <\omega_*$.
Also,  it is reasonable to
assume $\Delta z~ {\buildrel < \over \sim}~ r_1$.
Therefore,
$$
T_1\approx  \bar{\alpha}_1 r_1^3 B_1^2 \equiv
\bar{\alpha}~{\mu^2 \over
R_A^3}~,
\eqno(11)
$$
where $\bar{\alpha} \leq 1$ (the time average
of $\alpha(t)$) is a dimensionless
constant analogous to the $\alpha-$parameter
of Shakura (1973) and Shakura \& Sunyaev (1973).
(Note that because $r_1~ {\buildrel < \over \sim}~ R_A$,
$\bar{\alpha}$ and
$\bar{\alpha}_1 = (r_1/R_A)^3
\bar{\alpha}$ are of the same order of magnitude.)

Substituting equations (10) and (11) into
equation (9) gives
$$
{3\over 2}~(1-\delta^{3\over 2})~{\cal F}^{2\over 3}_o \left({r_A
\over R_A}\right)^{7\over 3} =
$$
$$
{\bar{\alpha}~r_A^{7\over 2} \over R_A^2~ r_{cr}^{3\over2}}
\left[1-\left({r_{cr}\over R_A}\right)^{3\over 2}
\delta^{3\over 2}\right] -{3\over 2}(1-\delta^{3\over 2})~.
\eqno(12)
$$
Here, we have introduced  two
characteristic radii -- the corotation radius,
$$
 r_{cr} \equiv \left( {GM \over \omega_*^2}\right)^{1 / 3}
\approx 1.5 \times 10^8 {\rm cm}~M_1^{1/3}
P_1^{2/3}~,
$$
with $P_1 \equiv (2\pi/\omega_*)/1{\rm s}$
the pulsar period and $M_1 \equiv M/M\odot$.
[For a young stellar object, $r_{cr} \approx
1.36\times 10^{12} M_1^{1/3} P_{10d}^{2/3}$,
where $P_{10d}$ is the period in units of
$10$ days.]
  The second is
the {\it nominal} Alfv\'en radius
$$
r_A \equiv \left[{ \mu^2 \over \dot{M}_{accr}
\sqrt{GM}}\right]^{2/ 7}
\approx 3.6\times 10^8{\rm cm}~{ \mu_{30}^{4/7}
\over \dot{M}_{17}^{2/7} M_1^{1/7}},
$$
where the accretion rate
$\dot{M}_{17}$ $ \equiv \dot{M}_{accr}/(10^{17} {\rm g/s})$
 with $10^{17}{\rm g/s}
\approx 1.6 \times 10^{-9} M_\odot/yr$ and $\mu_{30}
\equiv \mu/(10^{30} {\rm Gcm}^2)$ with
the magnetic field at the star's equatorial surface
$\mu/r^3 = 10^{12} G (10^6 {\rm cm}/r)^3$.
 [For a young stellar object $r_A \approx
1.81 \times 10^{12}{\rm cm}
 \mu_{36.5}^{4/7}/(\dot{M}^{2/7}M_1^{1/7})$,
where the normalization corresponds to a stellar
radius of $10^{11}$ cm, a surface magnetic
field of $3\times 10^3$ G, and an accretion
rate of $1.6 \times 10^{-8} M_\odot/{\rm yr}$.]
The corotation radius is the distance
from the star where the centrifugual force
on a particle corotating with the star ($\omega_*^2r$)
balances the gravitational attraction ($GM/r^2$).
The Alfv\'en radius $r_A$ is the
distance from a {\it non-rotating}
star where the free-fall of a quasi-spherical
accretion flow is stopped, which occurs
(approximately) where the kinetic
energy-density
of the flow equals the
energy-density of the star's
dipole field.
 Note that the assumptions leading
to equation (12) require $R_A > r_{cr}$.

Notice that $R_A$ (or $r_1 \sim R_A$)
is the {\it effective}
Alfv\'en radius for a {\it rotating} star.
It depends on {\it both} $r_A$ and $r_{cr}$
in contrast with the common notion that the
Alfv\'en radius is given by $r_A$
even for a rotating star.
>From equation (11), the {\it
spin-down} rate of the star is
$
I(d\omega_* / dt) =
- ~\bar{\alpha}~{\mu^2 /R_A^3}
$,
where $I$ is the moment of inertia of
the star (assumed constant).
Thus the spin-down rate
depends on both $r_A$ and $r_{cr}$.

 Figure 2 shows the dependence of $R_A$ on
$r_A$ and $r_{cr}$ for a sample case.
 For conditions
of a newly formed disk around a young pulsar,
the initial system point would be on the upper
left-hand part of the curve.
 Due to
the pulsar slowing down (assuming $\mu$ and
$\dot{M}$ constant), the system
point would move downward and to the right as indicated
by the arrow.
 In this region of the diagram, $R_A \approx
\sqrt{\bar{\alpha}}~r_A^{7/4}/r_{cr}^{3/4}
\propto \sqrt{\omega}$
(for $\delta \ll 1$), so that the
torque on the star is $T_1 \approx (GM\dot{M})^{3/2}
/(\mu \sqrt{\bar{\alpha}}~\omega^{3/2})$.  Thus
the braking index is $n=-3/2$, where
$n$ is defined by the relation
$\dot{\omega}_* = -{\rm const}~\omega_*^n$.
Numerically,
$${1\over P}{dP \over dt} =
{1.55 \times 10^{-5}\over {\rm yr }}~
{P_1^{5/2} (M_1~\dot{M}_{17})^{3/2} \over
\sqrt{\bar{\alpha}} ~\mu_{30}~ I_{45} }~,
\eqno(13)
$$
where $I_{45}\equiv I/(10^{45} {\rm g~cm}^2)$.
For $R_A \gg r_{cr}$, the mass accretion rate
to the star is  small
compared with $\dot{M}_{accr}$, but some
accretion may occur due
to `leakage' of relatively
low angular momentum plasma across field
lines near $r_1$ (Arons \& Lea 1976).

As $R_A$ decreases, the spin-down torque
on the star increases.  Over a long
interval, $R_A$ will decrease to a value
larger than $r_{cr}$ but not much larger.
In this limit, mass
accretion to the star $\dot{M}_1$ may
 become significant.  Our treatment
can be extended to this limit
by noting that $\dot{M}_1 =\dot{M}_2
-\dot{M}_{out}$, $\dot{L}_1=\dot{M}_1\omega_1r_1^2
-T_1$, and $\dot{E}_1=\dot{M}_1(-GM/2r_1)-\omega_*T_1$,
where $T_1$ is given by equation (11).
In this limit the accretion luminosity is
$GM\dot{M}_1/r_*$, where $r_*$
is the star's radius.   Figure 2 is not
changed appreciably for
$\dot{M}_1 < \dot{M}_{accr}$.

 Further spin-down of the star
will cause the system point in Figure 2
to approach the right-most part of the
curve.
 Further spin-down of the
star is impossible.
 At this point of the evolution,
the only possibility is a transition
to the spin-up regime.
 In this regime the effective
Alfv\'en radius is the `turnover radius'
$r_{to}$ of the disk
rotation curve calculated by
LRBK, the star spins-up at
the rate $I d \omega_*/dt = \dot{M}_{accr}
(GMr_{to})^{1/2}$, and most of the disk
accretion $\dot{M}_{accr}$ falls onto the
star.
 The dashed
horizontal line in Figure 2 indicates $r_{to}$
which is necessarily less than $r_{cr}$.

 The location of the turnover line $r_{to}$ in
Figure 2 suggests the possible evolution shown
by the sequence of points $a \rightarrow
b \rightarrow c\rightarrow d \rightarrow a$.
   The system can jump
down
from point $a$ where the star spins-down
to point $b$ where it spins-up.
 The spin-down torque at $a$ is
$-\bar{\alpha}\mu^2/R_{Aa}^3$
(where $R_{Aa}$ is the effective Alfv\'en
radius at point $a$), whereas
the spin-up torque at $b$ is
$\dot{M}_{accr} (GMr_{to})^{1/2}$.
 The magnitude
of the ratio of these torques is
$$
{{\rm spin\!-\!down} \over {\rm spin\!-\!up}} =
\bar{\alpha}\left({r_A \over R_{Aa}}\right)^{7/2}
\left({R_{Aa}\over r_{to}}\right)^{1/2}~.
\eqno(14)
$$
With the system on the $r_{to}$ line, it
evolves to the left.
  Because $r_{to} \leq r_{cr}$,
there must be
an upward jump from point $c$ to point $d$.
   For this case the torque ratio is given
by equation (14) with $R_{Aa} \rightarrow R_{Ad}$.
  From point $d$ the system evolves to the right.
For the example shown in Figure 2, the torque
ratio is $\approx 3.45$ for $a \rightarrow b$
whereas it is $\approx 0.87$ for $c \rightarrow d$.
   The vertical line $c\rightarrow d$ is at
the left-most position allowed
for the considered conditions, but
this transistion could also occur if the
line is shift to the right.
   The line $a \rightarrow b$  can be displaced
slightly to the right or it can be displaced
to the left to be coincident with the
$c \rightarrow d$ line.
   In the latter case the torque ratio for the
spin-down to spin-up jump is
approximately equal to the torque ratio for
the spin-up to spin-down jump and is $\approx 0.87$.
  The smaller the horizontal separation of the
$a \rightarrow b$ and the $c \rightarrow d$
lines, the shorter is the time interval
between jumps.

  Summarizing, we can say that the
horizontal locations of the transistions, $a \rightarrow b$,
and $c \rightarrow d$,
are indeterminate within a definite range.
  The locations of the jumps in the $(R_A,~r_{cr})$
plane may in fact be a
stochastic or chaotic in nature and give
 rise to chaotic hysteresis in the
of the spin-down/spin-up
behavior of the pulsar.
The jumps could be triggered by small variations in
the accretion flow ($\dot{M}$ for example)
and magnetic field configuration (the time-dependence
of $\alpha$ in the torque $T_1$).
  Analysis of  the accreting
neutron star system Her X-1
(Voges, Atmanspacher, \& Scheingraber 1987;
Morfill et al. 1989)
suggests that the intensity variations
are described by a low dimensional deterministic
chaotic model.
   The transistions between
spin-down and spin-up and the reverse transistions
may be described by an analogous model.

 The allowed values in Figure 2 have
$r_{cr}/r_A \leq {\rm const} \equiv k$,
where
$k \approx \bar{\alpha}^{2/3}$.
This
corresponds to pulsar periods
$$
P \leq 5.2{\rm s}
\left({\bar{\alpha}\over M_1^{5/7}}\right)
\left({\mu_{30}^2 \over \dot M_{17}}\right)^{3/7}~.
\eqno(15)
$$
For some long period pulsars such as GX 1+4
this inequality  points to  magnetic
moment values $\mu_{30}$ appreciably larger
than unity.
  Periods much longer than allowed by
(15) can result for pulsars which accrete
from a stellar wind (Bisnovatyi-Kogan 1991).
  [For a young stellar object, equation (15)
gives
$P \leq 8{\rm d}(\bar{\alpha}/M_1^{5/7})
(\mu_{36.5}^2/\dot{M}_{18})^{3/7}$.]

\section{Discussion}

This work presents a new investigation of
the `propeller' regime of disk accretion
to a rapidly rotating magnetized star.
The work considers the field configuration
proposed by LRBK, the theory of LBC on
magnetically driven jets, and the
conservation of mass, angular momentum,
and energy to derive an expression for the
effective Alfv\'en radius $R_A$ (equation 12)
and the spin-down torque
on the star $T_1$ (equation 11).
  Our work is in qualitative accord with that
of Li and Wickramasinghe (1997) who
also consider the propeller effect of Illarionov
and Sunyaev (1975).
  Our Figure 1 is similar to Figure 4 of
Li and Wickramasinghe for the spin-down regime,
and our earlier work on the spin-up regime
(LRBK) agrees with their Figure 3.
   We find that $R_A$ depends not only
on $\mu$, $\dot{M}$, and $M$, {\it but also}
on the star's rotation rate $\omega_*$.
   Because $R_A$ decreases as $\omega_*$
decreases, there is a
minimum value of $\omega_*$ or a maximum
value of the pulsar period $P=2\pi/\omega_*$.
   The model points to a mechanism
for `jumps' between spin-down and
spin-up evolution (and  the reverse
transition).  In our picture, in
a spin-down to spin-up transition,
for example, the effective
Alfv\'en radius decreases by an
appreciable factor going from
$> r_{cr}$ to $< r_{cr}$.
  The propeller goes from being ``on''
to being ``off'' in this transistion,
which is a change between two possible
equilibrium configurations.
  The transistions may be stochastic
or chaotic in nature with triggering
due to small variations in the
accretion flow or in the magnetic
field configuration.
  The ratio of the spin-down to
spin-up torques (or the ratio
for the reverse
transition) is found to be of
order unity (equation 14).
This agrees with observations of for
example Cen X-3 (Chakrabarty et al. 1993)
and GX 1+4 (Chakrabarty et al. 1997; Cui 1997).

\acknowledgements
{We thank Dr. Wei Cui for valuable
comments on this work.
This work was supported in part by
NSF grant AST-9320068.
Also, this work
was made possible in part by
Grant No. RP1-173 of the U.S.
Civilian R\&D Foundation for the
Independent States of the Former Soviet Union.
The work of
RVEL was also supported in part by NASA
grant NAG5 6311.}

\end{document}